\documentstyle[12pt,epsf]{article}

\def\be{\begin{equation}}
\def\ee{\end{equation}}
\def\bea{\begin{eqnarray}}
\def\eea{\end{eqnarray}}

\begin{document}

\begin{flushright}
hep-ph/0306150
\end{flushright}

\pagestyle{plain}

\def\e{{\rm e}}
\def\haf{{\frac{1}{2}}}
\def\tr{{\rm Tr\;}}
\def\goes{\rightarrow}
\def\ie{{\it i.e.}, }
\def\tcl{T_{\rm cl}}
\def\Goes {\Rightarrow}
\def\CX{{\cal X}}
\def\cm{{\rm c.m.}}
\def\fp{{\rm f.p.}}
\def\ca{{\cal A}}
\def\cf{{\cal F}}
\def\cd{{\cal D}}
\def\cv{{\cal V}}
\def\cvsym{{\cal V}_{{\rm sym.}}}
\def\cvnonsym{{\cal V}_{{\rm non-sym.}}}
\def\iphi{{\bf i}_\Phi}
\def\iv{{\bf i_v}}
\def\xgphi{{x\goes\Phi}}

\def\bsx{{\bf x}}
\def\bbx{{\bf X}}
\def\bsp{{\bf p}}
\def\bbp{{\bf P}}
\def\bba{{\bf A}}
\def\bsv{{\bf v}}
\def\bsk{{\bf k}}
\def\br{{\bf r}}

\begin{center}
\vspace{1.5cm}

{\Large {\bf Strings In Baryons And Matrix Coordinates}}

\vspace{.8cm}

Amir H. Fatollahi

\vspace{.5cm}

{\it Institute for Advanced Studies in Basic Sciences (IASBS),}\\
{\it P. O. Box 159, Zanjan 45195, Iran}

\vspace{.5cm}

{\sl fatho@mail.cern.ch}
\end{center}
\vskip .4 cm

\begin{abstract}
It is argued that the internal dynamics of a baryon, as a bound state of QCD-strings and quarks,
may be captured by a theory of matrix coordinates.
\end{abstract}

\vspace{.6cm}



\vspace{.6cm}


\newpage


There is a relatively common belief that gauge theories have some
kinds of string theory as their dual description. The supports for
this belief come from some phenomenological facts, such as Regge
trajectories, and also some theoretical considerations, like those
coming from large-$N$ and lattice gauge theories
\cite{tooft}\cite{wilson}. In the theoretical side, the
gauge/gravity duality in the AdS limit of brane backgrounds
\cite{9711200} has realized the idea by a definite relation
between the objects in two sides \cite{gkpw}. The loop-space
formulation \cite{loopspace1} is another example of efforts for
string theoretic presentation of gauge theory dynamics, still
exciting theoretical research \cite{loopspace2}.

In providing a dictionary between gauge and string theories, it is
usually assumed that the dynamics of the electric fluxes in
confined phase is presented by a string theory, probably in
non-critical dimensions. These electric fluxes are usually
referred under the names of ``flux-tubes" or ``QCD-strings." In
this picture, the QCD-strings are stretched between and are
responsible for the confinement of electric charges; similar to
the scenario we expect, via the Meisner effect, for the
confinement of a monopole and anti-monopole in a superconductor.
One justification for giving a role to these kinds of strings is
the linear potential that is usually supposed between the quark
and anti-quark of a mesonic state; a potential which can generate
Regge trajectories, and also is supported by lattice gauge theory
calculations \cite{wilson}.

Although the QCD-strings are usually considered for mesons, there
are quite reasonable proposals to introduce these strings to
baryons too. In fact the results by lattice calculations for the nature of
the inter-quark potential in a baryon strongly suggest that these kinds of
strings should be present in baryons. There are two main proposals
for the potential, the so-called: 1) $\Delta$-shape, and 2)
Y-shape. In the $\Delta$-shape ansatz, the total potential of
quarks is the sum of linear two-body ones, for three quarks as:
\bea
V(\br_1, \br_2, \br_3)=\frac{\sigma}{2} (r_{12}+
r_{23}+r_{31})
\eea
in which $r_{ij}=|\br_i-\br_j|$, and $\sigma$
is the tension of QCD-string in a mesonic state. The name $\Delta$
clearly comes from the proposed geometry by the potential, and
also the suggested form for the QCD-strings in a baryon. This
behavior is supported
by some lattice
calculations \cite{tsapalis}, at least for short
distances. In the Y-shape ansatz, the total potential is
introduced with the help of an extra point $\br_{\rm s}$, for
which the length $(r_{1{\rm s}}+r_{2{\rm s}}+r_{3{\rm s}})$ is
minimum ($r_{i{\rm s}}=|\br_i-\br_{\rm s}|$); the so-called
Steiner (or Fermat) point. Then the total potential for three
quarks is given by:
\bea
V(\br_1, \br_2, \br_3)= \sigma (r_{1{\rm
s}}+r_{2{\rm s}}+r_{3{\rm s}}).
\eea
The Y-shape ansatz has been used to extract the properties of
baryons \cite{isgur}, and also has found supports by the lattice
calculations \cite{takahashi}.
The field correlator method also produces results closer to
Y-ansatz \cite{simonov}.
Besides, the stringy shape of electric fluxes has
been revealed also by this method \cite{simonov}.

Due to geometrical reasonings \cite{bali}, the disagreement
between $\Delta$- and Y- ansatzes is at most about \%15, which is
still small to provide a final conclusion for lattice based
calculations. Also both proposals are almost equally suitable for
phenomenological purposes. One may state that by the present
results the potential approaches the $\Delta$-ansatz at short
distances, but rises like the Y-ansatz at large distances
\cite{edwards,bali}, and hence the $\Delta$ model is more
appropriate inside a hadron \cite{edwards,bali}. The deviations
from the Y-ansatz in small separations is interpreted in
\cite{simonov} by the depletion of the electric field at and
around the Steiner point (see Figs. by \cite{simonov}). The
depletion of electric field can be understood as the cancellation
of different components of electric fields when they arrive the
Steiner point. So by this interpretation we see that, quite
interestingly, there is a very preferred $\Delta$-shape for the profile of
electric field and also for the configuration of QCD-strings inside baryons,
specially for small separations of quarks.

The picture of baryons as bound states of quarks and QCD-strings
has been used for many years for phenomenological aims, and has
been able to produce a lots of considerable `numbers'. The natural
guess about the relevant theory is the formulation for the system
of ``three point-like masses bounded by relativistic strings"
\cite{isgur} \cite{stringbaryon}. By this formulation one can
explain somehow directly some known facts, such as Regge behavior.
In spite of long history of the efforts based on string-quark
picture, however, there are difficulties with this formulation.
One is that the string theory in use is
in non-critical dimensions, and so in principle, one expects that
in the level of complete quantum theory some kinds of anomalies or
infinities appear. The source of anomalies or infinities simply is
that, after all, we are dealing with a field theory, living on the
world-sheets of some strings. The other problem is related to high
non-linearity of the system; a non-linearity which can not be get
rid of easily due to non-critical dimension.

As the existence of QCD-strings is well accepted, and as the models based
on quark-string picture have shown to be practically useful, it is apposite
to suggest models that while they capture the essential features of the
quark-string picture, they avoid difficulties that one usually is faced
in studying the highly non-linear field theory on the world-sheet.
To avoid unwanted infinities one way is dealing with a quantum mechanical
system rather than a quantum field theory. In
this direction, one analogy with the picture we imagined for
baryons and the introduction of D0-branes in string theory may be
suggestive. D0-branes are defined as massive point-like objects to whom
the open strings end \cite{9510017}. While the open strings are stretched between the
D0-branes, they join D0-branes together, making a bound state
of D0-branes and strings. Here two cases can be considered, as one
assumes the open strings are oriented or unoriented. Eventually it
appears that in a bound state of $N$ D0-branes the relevant
degrees of freedom in each direction of space, rather than $N$,
are $N^2$ in the case of oriented strings, and $N(N+1)/2$ for
unoriented case. These degrees of freedom may be represented by
matrices belonging to U($N$) and O($N$) algebras, for oriented and
unoriented cases, respectively. So one may argue that the degrees
of freedom for system of $N$ D0-branes, rather than $d.N$ numbers
in $d$ dimensions, are $d$ matrices of dimension $N\times N$
\cite{9510135}\cite{tasi}. The matrix coordinates find this
interpretation that the diagonal elements capture the dynamics of
D0-branes, and the dynamics (oscillations) of strings are encoded
in off-diagonal elements; for example, the dynamics of the
string(s) stretched between $a$-th and $b$-th D0-branes has (have)
been encoded in the element $\bbx_{ab}\;(a\neq b)$.

In \cite{02414,fat021} the action for D0-branes was considered as
a model for QCD purposes. The first motivations for these studies
originated by some early results which appear quite suggestive to
take serious analogy between quark-strings systems and bound state
of D0-branes. For this one should assume that the theory for
matrix coordinates, perhaps coming from a critical string theory,
should be also suitable for non-critical dimensions. This needs
justification, though since the theory in its quantized form is
just a quantum mechanics, one might be so hopeful that the results
should be free from very bad and unwanted behaviors. The concerned
model in \cite{02414,fat021} has shown its ability to reproduce
and recover some features and expectations in hadron physics. Some
of these features and expectations are: the linear inter-quark
potentials, the behavior of total scattering amplitudes, rich
polology of scattering amplitude, behavior in large-$N$ limit, and
the whiteness of baryons with respect to the SU($N$) sector of the
external fields. The purpose of this Letter is to restate and
refresh the main idea, in particular, via appreciating the
suggestive and insightful presence of QCD-strings in a baryon, revealed by the
recent lattice calculations and the field correlator method.

The dynamics for the matrix coordinates of D0-branes is formulated
as a matrix theory. Although the exact form of the action should
come from stringy calculations, based on some general arguments
one suggestion is:
\bea\label{action5}
S[\bbx,
a_t]=\int dt \;\tr \bigg(\frac{1}{2} m D_t\bbx\cdot D_t\bbx
+qD_t\bbx\cdot \bba(\bbx,t)\nonumber\\
 - qA_0(\bbx,t) +\frac{m}{4l^4}[X^i,X^j]^2+\cdots)\bigg),
\eea
where $\tr$acts on the matrix structure, and
$D_t\bbx=\dot{\bbx}+i[a_t,\bbx]$ is the covariant velocity, with
$a_t(t)$ as the one dimensional $N\times N$ gauge field. $l$ in the
language of string theory is order of the string
length. The potentials ($A_0(\bbx,t), \bba(\bbx,t)$) are
functionals of symmetrized products of the matrix coordinates, and
``$\cdots$" is for $O(\bbx^6)$ and higher non-symmetrized terms,
and also non-linear terms in velocity $D_t\bbx$. We note that the
fields $(A_0(\bbx,t),\bba(\bbx,t))$ appear as $N\times N$
hermitian matrices due to their functional dependence on the
matrix coordinate $\bbx$. One can check easily that action
(\ref{action5}) is invariant under the symmetry transformations
\cite{0103262,0104210,0108198}:
\bea\label{NAT}
\bbx&\goes& \bbx'=U \bbx U^\dagger,\nonumber\\
a_t&\goes& a'_t=U a_t U^\dagger -i
U \frac{d}{dt}U^\dagger,\nonumber\\
A_i(\bbx,t)&\goes& A'_i(\bbx',t)=
U A_i(\bbx,t)U^\dagger+iU\delta_i U^\dagger,\nonumber\\
A_0(\bbx,t)&\goes&  A'_0(\bbx',t)= U
A_0(\bbx,t)U^\dagger-iU\partial_t U^\dagger,
\eea
where $U\equiv U(\bbx,t)=\exp(i\Lambda)$ is arbitrary up to the condition that
$\Lambda(\bbx,t)$ is totally symmetrized in the $X^i$'s. In above
$\delta_i$ is the functional derivative $\frac{\delta} {\delta
X^i}$. We recall that in approving the invariance of the action,
the symmetrization prescription on the matrix coordinates plays an
essential role \cite{0103262,0104210}. The above transformations
on the gauge potentials are similar to those of non-Abelian gauge
theories, and we mention that it is just the consequence of
enhancement of degrees of freedom from numbers to matrices. In
other words, we are faced with a situation in which `the rotation
of fields' is generated by `the rotation of coordinates'
\cite{0104210}. Despite the non-Abelian behavior of the gauge
transformations, since they are defined by just one function
$\Lambda(\bsx,t)$ after replacing ordinary coordinates by their
matrix partners, {\it i.e.}  $\bsx\goes\bbx$, the transformations
are not equivalent to non-Abelian ones \cite{03115}. After all, it
is quite natural to interpret the fields $(A_0,\bba)$ as the
external gauge fields that the constituents, whose degrees of
freedom are included in the matrix coordinate, interact with them
\cite{0108198}.

By ignoring the external fields $(A_0,\bba)$, one can find the
effective theory for the matrices \bea
\bbx(t)&=&{\rm diag.}\; (\bsx_1(t),\cdots, \bsx_N(t)),\nonumber\\
a_t(t)&=&{\rm diag.} \;(a_{t1}(t),\cdots,a_{tN}(t)), \eea with
$\bsx_a=\bsx_{a0}+\bsv_a t$, $a=1,\cdots,N$. This configuration
solves the equations of motion, and describes the ``classical"
free motion of $N$ D0-branes, neglecting the effects of the
strings. The situation is different when we consider quantum
effects, and it will be realized that the dynamics of the
off-diagonal elements affect the dynamics of D0-branes
significantly. Concerning the effect of the strings, one may try
to extract the effective theory for D0-branes. In particular, it
will be found out that the commutator potential is responsible for
the formation of the bound state, and by a simple dimensional
analysis we understand that the size of the bound state, $\ell$,
is $\sim m^{-1/3} l^{2/3}$. For the static configuration
\cite{02414} one can easily calculate one-loop effective potential
between the D0-branes, getting \cite{fat021,02414}:
\bea\label{one-loop}
V_{{\rm one-loop}}\sim\sum_{a>b=1}^N \frac{|\bsx_a-\bsx_b|}{l^2}.
\eea
This result shows that the matrix theory, up to this order of
expansion in use, is in accordance with $\Delta$-ansatz for
potential. Since the original theory is invariant under the
rotation among the indices $1,\cdots, N$, only the states which
are singlets under the global rotation among the indices can be
accepted as the physical states of effective theory for diagonal
elements.

The concerned model above is in the non-relativistic limit, and though
it may appear suitable for heavy quarks, for light or massless
quarks we should change our approach. The formulation in use is that of
the M(atrix) model conjecture \cite{9610043}, accompanied with the
gauge field terms. To approach the covariant formulation,
following finite-$N$ interpretation of \cite{9704080}, it is
reasonable to interpret things in the DLCQ (Discrete Light Cone
Quantization) framework, see \cite{0108198} and Appendix of \cite{fat021}.

In \cite {03115} a construction of matrix coordinates from some
basic tools and expectations in QCD was proposed. In particular,
by giving an equivalent role to color labels in gauge theory, and
the so-called Chan-Paton labels in open strings, one observes that
by the wave functions of individual quarks, say
in a Hartree-Fock approximation for the internal dynamics of
baryon, one can define the matrix coordinates by their elements:
\bea
\bbx_{ab}(t)&=&\langle\psi_b(t) |\hat{\bsx}|\psi_a(t)\rangle\nonumber\\
&=&\int d\bsx\; \psi^*_a(\bsx,t) \bsx \psi_b(\bsx,t)
=\bbx^*_{ba}(t).
\eea
It was argued in \cite{03115} that the amount of simultaneous diagonalizability
of matrix coordinates, and their generalizations to higher moments, can be used
as a criteria for identifying the confined phase of a gauge theory.

One may wonder why a formulation of a non-Abelian gauge theory in
confined phase, according to a proposal for covering the stringy
behavior, should be involved by matrix coordinates. In
\cite{0104210,fat021} a conceptual relation between appearance of
matrix coordinate in formulation of a non-Abelian gauge theory and
one interpretation of special relativity is mentioned. According
to an interpretation of the special relativity, it is meaningful
if the `coordinates' and the `fields' that are involved in a theory have
some kinds of similar characters. Accordingly, we see that
both the space-time coordinates $x^\mu$ and the electro-magnetic
potentials $A^\mu(x)$ transform as a four-vector under the boost
transformations. Also by this way of
interpretation, the superspace formulations of supersymmetric
field and superstring theories are the continuation of the
special relativity program: adding spin-half coordinates as
representatives of the fermionic degrees freedom of the theory. It
may be argued that the relation between `matrix coordinates' and
`matrix fields' is one of the expectations which is supported by
the spirit of the special relativity. By previous discussion
we recall, 1) the matrix character of gauge fields is the result
of dependence of them on matrix coordinates \cite{0104210}, 2) the
symmetry transformations of gauge fields are induced by the
transformations of matrix coordinates \cite{0104210}, 3) the
transformations of fields in the theory on matrix space appeared
to be similar to those of non-Abelian gauge theories.
We note that by the picture supposed in this work, we expect that
the matrix coordinates become relevant just for studying the internal dynamics of
bound states, a situation reffered in \cite{0103262,0104210,0108198,03115}
as `confined non-commutativity.'
This way of interpretation may lead us to conclude that
the non-Abelian gauge fields in a confined theory do not have an
independent character, and they are introduced to the formalism
just through the functional dependence of Abelian gauge fields
on the matrix coordinates of `bounded quarks'. It is very interesting
when we note that by the present status of the experimental data, the
existence of pure gluonic states, the so-called glueballs, is quite
doubtful. This lack of detection, while it is quite unexpected by the
masses we expect for glueballs, may be taken as a support for this
interpretation.

\vspace{.5cm}
{\bf Acknowledgments}
The author is thankful for useful comments by M. Khorrami, M. M. Sheikh-Jabbri and
S. Parvizi.

\end{document}